\documentclass[conference]{IEEEtran}

\usepackage{amsmath,epsfig}
\usepackage{bm}
\usepackage{latexsym}
\usepackage{amsmath}
\usepackage{amssymb}
\usepackage{graphicx}

\def\bs{\mathbf{s}}
\def\bv{\mathbf{v}}
\def\bw{\mathbf{w}}
\def\bx{\mathbf{x}}
\def\by{\mathbf{y}}

\def\br{\mathbf{r}}
\def\bpsi{\bm{\psi}}

\begin{document}

\title{A High-Throughput Cross-Layer Scheme for Distributed
Wireless Ad Hoc Networks}

\author{\authorblockN{Athina P. Petropulu, Lun Dong}
\authorblockA{Department of Electrical and Computer Engineering\\
Drexel University, Philadelphia, PA 19104} \and
\authorblockN{H. Vincent Poor}
\authorblockA{School of Engineering and Applied Science\\
Princeton University, Princeton, NJ 08544} } \maketitle
\footnote{This work was supported by the National Science Foundation
under Grants ANI-03-38807, CNS-06-25637 and CNS-04-35052, and by
the Office of Naval Research under Grant ONR-N-00014-07-1-0500}
{\small
\begin{abstract}
 In wireless ad hoc networks, distributed nodes can
collaboratively form an antenna array for long-distance
communications to achieve high energy efficiency. In recent work,
Ochiai, et al., have shown that such collaborative beamforming can
achieve a
statistically nice beampattern with a narrow main lobe and low side
lobes. However, the process of collaboration introduces significant
delay, since all collaborating nodes need access to the same
information. In this paper, a technique that significantly
reduces the collaboration overhead is proposed.
 It consists of two phases. In
the first phase, nodes transmit locally in a random access fashion.
Collisions, when they occur, are viewed as linear mixtures of the
collided packets. In the second phase, a set of cooperating nodes
acts as a distributed antenna system and beamform the received analog
waveform to one or more faraway destinations. This step requires
multiplication of the received analog waveform by a complex number,
which is independently computed by each cooperating node, and which
enables separation of the collided packets based on their final
destination. The scheme requires that each node has global knowledge
of the network coordinates. The proposed scheme can achieve high
throughput, which in certain cases exceeds one.
\end{abstract}}
\textbf{Keywords}: distributed wireless systems, cooperation,
beamforming\\

\section{Introduction}
 Energy is often
a scarce commodity in wireless sensor networks, as wireless sensors
typically operate on batteries, which in many cases are hard to
replace. Similarly, due to cost consideration, nodes in wireless ad
hoc sensor networks are commonly equipped with only a single
omnidirectional antenna. Thus, in order to transmit information over
long distances while conserving energy and  maintaining a certain
transmission power threshold,
 multihop networks have been the preferred solution. However, there are several challenges
in transmitting real-time services over multiple hops. For example,
the traditional CSMA/CA based media access control for avoiding
collisions does
not work well in a multihop scenario because transmitters are often
 out of
reach of other users' sensing range. Thus, as packets travel across
the network, they experience interference and a large number of
collisions, which introduces  delays. Also, multihop networks
require a high node density  which makes routing more difficult
and affects the reliability of links \cite{Gharavi}.

Recently, a collaborative beamforming technique was proposed in
\cite{Ochiai}, in which randomly distributed nodes in a network
cluster form an antenna array and beamform data to a faraway
destination without each node exceeding its power constraint. The
destination receives data with high signal power. Beamforming with
antenna arrays is a well studied technology; it provides
space-division multiple access (SDMA) which enables significant
communication rate increase. The main challenge with implementing
beamforming based on randomly distributed nodes in the network is
that the geometry of the network changes dynamically. It was shown
in \cite{Ochiai} that such a distributed antenna array can achieve a
  nice average beampattern with a narrow main lobe and low side
lobes. The directivity of the pattern increases as the number of
collaborating nodes increases. Such an approach, when applied in the
context of a multihop network reduces the number of hops needed,
thereby reducing packet delays and improving throughput. However,
one must take into account the overhead required for node
collaboration, i.e., sharing of the information to be transmitted jointly
by the nodes. If a time-division multiple-access (TDMA) scheme
were to be employed, the information-sharing time would increase
proportionally to the number of nodes involved in the collaboration.

In this paper we present a technique based on the idea of
collaborative beamforming, that reduces the time required for
information sharing. The technique also allows different nodes in
the network to transmit simultaneously. Collaborating nodes receive
linear mixtures of the transmitted packets. Subsequently, each
collaborating node transmits a weighted version of its received
signal. The weights are such that one or multiple beams are formed,
each focusing on one destination node, and reinforcing the signal
intended for a particular destination as compared to the other
signals. Each collaborating node computes its weight based on the
channel coefficients between sources and itself, estimated by
orthogonal node IDs embedded in the packets. The proposed scheme
achieves higher throughput and lower delay with the cost of lower
SINR as compared to \cite{Ochiai}.

\section{Background on collaborative beamforming} \label{back}
For simplicity, let us assume that the sources and destinations are
coplanar. We index source nodes using a subscript $i,$ with
$t_i$ denoting the $i$-th node. The locations of these nodes
follow a uniform distribution
over a disk of radius $R,$ and it is assumed that each node knows
its own location. We denote the location of $t_i$
in polar coordinates with respect to the center of the disk by
$(r_i,\psi_i).$

Suppose that a set of $N$ nodes designated as collaborating nodes
have access to the same signal, i.e., $s(n)$, whose destination is
at azimuthal angle  $\phi_0$. Let $d_i(\phi_0)$ denote the distance between
the $i$-th collaborating node and the destination. The initial phases
at the collaborating nodes are set to
\begin{equation}
\Psi_i(\phi_0)=-\frac{2\pi}{\lambda} d_i(\phi_0), \ i=1,...,N
\label{Psi1}
\end{equation}
which requires knowledge of distances relative to wavelength between
nodes and destination, and applies to the closed-loop case
\cite{Ochiai}. Alternatively, the initial phase of node $i$  could be:
\begin{equation}
 \Psi_i(\phi_0)=\frac{2\pi}{\lambda} r_i \cos(\phi_0-\psi_i) \label{Psi2} \end{equation}
which requires knowledge of the node's position relative to some common
reference point, and corresponds to the open-loop case
 \cite{Ochiai}. In both cases synchronization is needed, which can
 be achieved via the use of the Global Positioning System (GPS).

The channels between collaborating nodes and destination are assumed
to be idential for all nodes. The corresponding array factor given
the collaborating nodes at radial coordinates ${\br}=[r_1,...,r_N]$  and
azimuthal coordinates ${\bpsi}=[\psi_1,...,\psi_N]$ is
\begin{eqnarray}
F(\phi|{\bf r,\bpsi})&=& \frac{1}{N}\sum_{i=1}^N e^{j\Psi_i(\phi_0)}
e^{j\frac{2\pi}{\lambda}d_i(\phi)}
\end{eqnarray}

Under far-field assumptions, the array factor becomes \cite{Ochiai}:
\begin{eqnarray} \label{100}
F(\phi|{\bf r,\bpsi})&=&\frac{1}{N}\sum_{i=1}^N  e^{j
\alpha(\phi;\phi_0)z_i} \label{2}
\end{eqnarray}
where  $\alpha(\phi;\phi_0)=4\pi\tilde R
\sin(\frac12(\phi_0-\phi))$, $\tilde R=R/\lambda$  and
$z_i=(r_i/R)\sin(\psi_i-\frac12(\phi_0+\phi))$ with the following
pdf:
\begin{eqnarray}
f_{z_i}(z)=\frac{2}{\pi} \sqrt{1-z^2}, & \ -1\le z\le 1
\end{eqnarray}
Finally, the average beampattern can be expressed as \cite{Ochiai}
\begin{eqnarray} \label{101}
P_{\mathrm{av}}(\phi)&=& E_z\{ |F(\phi|{\bf z})|^2 \} \nonumber\\
&=& \frac{1}{N}+\left(1-\frac{1}{N}\right)\left|
2\frac{J_1(\alpha(\phi;\phi_0) )}{\alpha(\phi;\phi_0)} \right|^2
\end{eqnarray}
where $J_1(.)$ is the first-order Bessel function of the first kind.
When plotted as a function of $\phi$, $P_{\mathrm{av}}(\phi)$
exhibits a main lobe around $\phi_0$, and side lobes away from
$\phi_0$. It equals one in the target direction, and the sidelobe
level approaches $1/N$ as the angle moves away from the target
direction. The statistical properties of the beampattern were
analyzed in \cite{Ochiai}, where it was shown that under ideal
channel and system assumptions, directivity of order $N$ can be
achieved asymptotically with $N$ sparsely distributed sensors.

As we have noted, all of the collaborating nodes must have the same
information to implement beamforming. Thus, the active nodes need to
share their information symbols with all collaborating nodes in advance.
If a time-division multiple-access (TDMA) scheme were to be
employed, the information-sharing time would increase proportionally
to the number of active nodes. In the following, we  propose a novel
scheme to reduce the information-sharing time and also allow nodes
in the network to transmit simultaneously.

\section{The proposed scheme} \label{proposed}

Here we refine the model of \cite{Ochiai}, focusing more directly on the
physical models for the signal, fading channel
and noise. Besides the assumptions in Section \ref{back}, we will
further assume the followings:

(1) The network is divided into clusters, so that nodes in a cluster
can hear each other. In each cluster there is a node designated as
the cluster-head (CH). Nodes in a cluster do not need to transmit
their packets through the CH.

(2) A slotted packet system is considered, in which each packet
requires one slot for its transmission. Perfect synchronization is
assumed between nodes in the same cluster.

(3) Nodes operate under half-duplex mode, i.e., they cannot receive
while they are transmitting.

(4) Nodes transmit packets consisting of PSK symbols having the same
variance $\sigma_s^2$. Each transmitted packet contains (in fixed
locations) a set of pilots comprising the user ID, followed by a set
of pilots comprising the destination information.

(5) Communication takes place over flat fading channels. The channel
gain during slot $n$ between source $t_i$ and collaborating node
$c_j$ is denoted by $a_{ij}(n)$. It does not change within one slot,
but can change between slots.  The gains
$a_{ij}$ follows a Rayleigh fading model, being
 i.i.d. complex Gaussian random variables with
zero means and variances $\sigma_a^2$.

(6) The complex baseband-equivalent channel gain between nodes
$c_i,d_j$ is \cite{tse-book}: $ b_{ij}= b \cdot e^{{j\frac{2\pi
r_{ij}}{\lambda}}} $ where $r_{ij}$ is the distance between nodes
$c_i,d_j,$ and  where $b$ is the path loss. The distances between
collaborating nodes and destinations are much greater than distances
between source and collaborating nodes. Thus, $b$ is assumed to be
identical for all collaborating nodes and equals the path loss
between the origin of the disk over which the nodes are distributed
to the destination.

(7) the noise vectors are uncorrelated, complex, zero-mean white Gaussian
vectors.

Suppose that cluster $C$ contain $J$ nodes. At slot $n$, nodes
$t_1,\ldots, t_K$ need to communicate with node $d_1,\ldots,d_K$
that belong to cluster $C_1,\ldots,C_K$, respectively. The azimuthal
angle
of destination $d_i$ is denoted by $\phi_i$. The packet
transmitted by $t_j$ consists of $M$ symbols $\bs_j(n) \triangleq
[s_j(n;0), \ldots, s_j(n;M-1)]$. Due to the broadcast nature of the
wireless channel, non-active nodes in cluster $C$ hear a collision,
i.e.,  a linear combination of the transmitted symbols. More
specifically, node $c_i$ hears the signal
\begin{equation} \label{crec}
\bx_i(n)=\sum_{j=1}^K a_{ji}(n)\bs_j(n) + \bw_i(n)
\end{equation}
where $\bw_i(n)=[w_i(n;0),\ldots, w_i(n;M-1)]$ is noise vector with
 variance $\sigma_w^2$ at the receiving node $c_i$.

Once the CH establishes that there has been a transmission, it
initiates a collaborative transmission period (CTP), by sending to
all nodes a control bit, e.g., $1$, via an error-free control
channel. The CH will continue sending a $1$ in the beginning of each
subsequent slot until the CTP has been completed. The cluster nodes
cannot transmit new packets until the CTP is over.

Suppose that each transmitted packet includes an ID sequence, so
that IDs are orthogonal between different users. The channel
coefficients can be estimated by cross correlating $\bx_i(n)$ with
known user IDs as in \cite{alliances}. If the magnitude of the
cross-correlation in greater than some threshold, then the
corresponding user is in the mixture, and the value of the
cross-correlation provides the corresponding channel coefficient.
The information of destination nodes could be obtained in a similar
way.

Each node $c_i$ uses cross-correlation operations with known
orthogonal user IDs to determine which users are in the mixture, the
corresponding destinations, and also the coefficients $a_{ij}$.

Let $d_m$ denote the destination of $\bs_m(n)$. In slot $n+m, \
m=1,\ldots ,K$, each collaborating node $c_i$ transmits the signal:
\begin{equation}
 \tilde \bx_i(n+m)=\bx_i(n) \mu_m a^*_{mi}(n) e^{-j
 \frac{2\pi}{\lambda}d_i(\phi_m)}
\end{equation}
where $d_i(\phi_m)$ denotes the distance between user $i$ and
destination node $d_m$ with azimuth $\phi_m$, and $\mu_m$ is a
scalar to adjust transmit power and same for all collaborating
nodes. In addition, $\mu_m$ is of the order of $1/N$.

Given the collaborating nodes at radial coordinates ${\br}=[r_1,...,r_N]$,
azimuthal coordinates
 ${\bpsi}=[\psi_1,...,\psi_N]$ and the path loss $b$, the
received signal at direction $\phi$ is:
\begin{equation} \label{rec}
\by(\phi;m|{\br,\bpsi})= \sum_{i=1}^N b \tilde \bx_i(n+m) e^{j
\frac{2\pi}{\lambda}d_i(\phi)} + \bv(n+m)
\end{equation}
where $\bv(n+m)$ is the noise vector with
 variance $\sigma_v^2$ at the receiver during
slot $n+m$.

Let us consider the received signal at the destination $d_m$ at
angle $\phi_m$ during slot $n+m$: {\setlength\arraycolsep{0.1em}
\begin{eqnarray}
\by(\phi_m;m)&=& \sum_{i=1}^N b \tilde \bx_i(n+m) + \bv_m(n+m)\nonumber \\
& =& \mu_m b \sum_{i=1}^N [|a_{mi}(n)|^2 \bs_m(n) +a^*_{mi}(n) \bw_i(n) \nonumber\\
&& +a^*_{mi}(n) \sum_{j=1 \ \atop j \ne m }^K a_{ji}(n)\bs_j(n)] +
\bv_m(n+m)
\end{eqnarray}}

As  $N\rightarrow \infty$, $\frac{1}{N}\sum_{i=1}^N |a_{mi}(n)|^2
\rightarrow E\{ |a_{mi}|^2\}=\sigma_a^2 $. Also,  $\frac{1}{N}
\sum_{i=1}^N \ a^*_{mi}(n)a_{ji}(n) \rightarrow E\{
a_{mi}^*a_{ji}\}=0$, due to the fact that for $j\ne m$, the channel
coefficients $a_{mi}, a_{ji}$ are uncorrelated and have zero mean.
Finally, as $N\rightarrow \infty$ and omitting the noise,
$\by(\phi_m;m|{\bf r,\bpsi}) \rightarrow N \mu_m b_m \sigma_a^2
\bs_m(n)$. Thus, the destination node $d_m$ receives a scaled
version of $\bs_m(n)$. The beamforming step is completed in $K$
slots, reinforcing one source signal at a time.

Assuming that all of  the $K$ source packets have distinct destinations
at different resolvable directions, multiple beams can be formed in
one slot, each beam focusing on one direction and reinforcing one
source signal. The transmitted signal from each collaborating node
would be:
\begin{equation} \label{trans}
 \tilde \bx_i(n+1)=  \bx_i(n) \sum_{m=1}^K \mu_m a^*_{mi}(n) e^{-j \frac{2\pi}{\lambda}d_i(\phi_m)}
\end{equation}
The received signal at destinations would then be:
\begin{eqnarray} \label{mulrec}
\by(\phi|{\br,\bpsi})= b \tilde \bx_i(n+1)
e^{j\frac{2\pi}{\lambda}d_i(\phi)} +\bv(n+1)
\end{eqnarray}
It can be shown that as $N\rightarrow \infty$ and omitting noise,
$\by(\phi_{m}|{\br,\bpsi})\rightarrow N \mu_m b \sigma_a^2
\bs_{m}(n)$, for $m=1,...,K$. Thus, each of the $K$ beams transmits
a scaled version of a source signal to its destination. Based on
(\ref{rec}), (\ref{mulrec}), we can easily extend the mathematical
formulation to the scenario that not all of the $K$ packets have
distinct destinations.

In the rest of the paper, for simplicity we will consider only the
case in which a single beam is formed during slot $n+m$, focusing on
destination $d_m$. The results  obtained  under this assumption
can be readily extended to
multiple simultaneous beams. The time index can be omitted without
causing confusion. Also, we are particularly interested in the
average properties over a statistical ensemble,
 so the analysis can be based on one
sample of a packet assuming samples are independent. Substituting
$\by$, $\tilde{\bx}$, $\bx$, $\bw$, $\bv$ in
(\ref{crec})-(\ref{rec}) by $y$, $\tilde{x}$, $x$, $w$, $v$ (i.e.,
with one of their samples) respectively, we get:
\begin{equation} \label{recnew}
y(\phi;m|{\br,\bpsi})= \sum_{i=1}^N b \tilde x_i e^{j
\frac{2\pi}{\lambda}d_i(\phi)} + v
\end{equation}

\section{Performance of the average beampattern} \label{avgbeam}
\label{sec:avgbeam} In this section we analyze the average
beampattern.  Under the far-field assumption, and following the
steps in \cite{Ochiai}, (\ref{recnew}) can be expressed as:
\begin{equation} \label{beam1}
y(\phi; m|{\br,\bpsi})= \sum_{i=1}^N \mu_m b a_{mi}^* e^{-j
\alpha(\phi;\phi_m)z_i} ( \sum_{j=1}^{K}{a_{ji} s_j}+w_i )+ v
\end{equation}
where $\alpha(\phi;\phi_m)$ and $z_i$ are the same as those in
(\ref{100}).

The far-field beampattern or the received power is defined as:
\begin{equation}
P(\phi)=|y(\phi; m|{\br,\bpsi})|^2 \end{equation}

Taking into account the assumptions on the channel coefficients, it
can be readily shown that the average beampattern equals: {
\setlength\arraycolsep{0.1em}
\begin{eqnarray}
P_{\mathrm{av}}(\phi)& \triangleq  &
E_{z,a,w,v}\{P(\phi;\phi_m)\}\nonumber\\
 &=& \mu_m^2 b^2E\{ |s_m|^2
\sum_{i=1}^N |a_{mi}|^4 \nonumber\\
&&+ |s_m|^2 \sum_{i=1}^N \sum_{\ell=1 \atop \ell\ne i}^N
|a_{mi}|^2|a_{m\ell}|^2
e^{-j\alpha(\phi;\phi_m)(z_i-z_{\ell})}\nonumber \\
& & +\sum_{j=1 \atop j \neq m}^{K} |s_j|^2 \sum_{i=1}^{N}|a_{mi}|^2
|a_{ji}|^2 \nonumber + \sum_{i=1}^N {|a_{mi}|^2}|w_i|^2 +|v|^2 \} \nonumber \\
&=&N \mu_m^2 b^2 [ 2\sigma_s^2 \sigma_a^4 + (N-1)\sigma_s^2
\sigma_a^4 E\{ e^{-j\alpha(\phi;\phi_m)(z_i-z_{\ell})} \}
\nonumber\\
&&+ (K-1)\sigma_s^2 \sigma_a^4 +\sigma_w^2 \sigma_a^2] + \sigma_v ^2
\end{eqnarray}}
Then,
\begin{equation} \label{avgbeampattern}
P_{\mathrm{av}}(\phi) = N^2 \mu_m^2 b^2 \sigma_s^2 \sigma_a^4
\left(\frac{\beta}{N} + (1-\frac{1}N )\left|2\cdot
\frac{J_1(\alpha(\phi;\phi_m))}{\alpha(\phi;\phi_m)}
\right|^2\right)
\end{equation}
where
\begin{equation}
\beta=K+1+\frac{\sigma_w^2}{\sigma_s^2\sigma_a^2}+\frac{N\sigma_v^2}{\mu_m^2
b^2 \sigma_s^2\sigma_a^4}
\end{equation}

The term $N\mu_m^2 b^2 \sigma_s^2 \sigma_a^4 \beta$ represents the
average power of the sidelobes that is independent of the angle
$\phi$. Note that (\ref{avgbeampattern}) is of the similar form with
(\ref{101}). Thus, other properties of the average beampattern like
peak/zero positions and 3-dB bandwidth/sidelobe region can be easily
obtained based on corresponding results of \cite{Ochiai}, and are
omitted here.

\section{Performance of the network}
\label{network}

\subsection{Throughput}

Suppose that $K$ packets need to be transmitted. For the
collaborative beamforming scheme of \cite{Ochiai}, each packet  must
be shared by the $N$ beamforming nodes. If one wished to avoid
collisions, TDMA would be a natural way to implement information
sharing. With the use of TDMA, in each slot one active node is
scheduled to broadcast its packet to other nodes within the same
cluster. The sharing of information would require $K$ slots. Via the
use of multiple beams, the beamforming to the destination would
require $1$ slot if destinations are distinct, or up to $K$ slots in
the worst case where all nodes have the same destination. Thus, for
the scheme of \cite{Ochiai}, the throughput, $T$ satisfies: $1/2 \le
T \le K/(K+1)$.

In the proposed scheme, combinations of $K$ packets enter the system
and reach $N$ collaborating nodes in $1$ slot. Via multiple beams,
the packets can be delivered to their destinations in one additional
slot, if the destinations are distinct. The throughput is then
$K/(1+1)=K/2$. If two or more packets have a common destination, the
beamforming will need to take more than one slot. In the worst case
where all packets have the same destination, $K$ slots will be
needed, resulting in a total throughput of $K/(1+K)$. Thus, $K/(1+K)
\le T_{\mathrm{proposed}}\le K/2$. Note that the throughput of the
proposed scheme could be greater than 1.

\subsection{Transmit Power and Average SINR}
In this section, we analyze the SINR under the same transmit power for the
scheme of \cite{Ochiai} and the proposed scheme.

Although the focus in \cite{Ochiai} was on the statistical properties
of the  beampattern, we can extend (\ref{100})-(\ref{101}) to a
physical model including signal, path loss and noise. The received
signal of the destination $d_m$ is given by:
\begin{equation} \label{201}
\overline{y}(\phi; m|{\br,\bpsi})= \sum_{i=1}^N b \bar{\mu}_m s_m
e^{-j \alpha(\phi;\phi_m)z_i}+ v
\end{equation}

$c_i$ simply transmits $\bar{\mu}_m s_m e^{-j \frac{2\pi}{\lambda}
d_i(\phi_m)}$, and the transmit power is thus $\bar{\mu}_m^2
\sigma_s^2$. The average beampattern at the target direction
$\phi_m$ is:
\begin{equation} \label{202}
\overline{P}_{\mathrm{av}}(\phi_m)=N^2 \bar{\mu}_m^2
b^2\sigma_s^2+\sigma_v^2
\end{equation}
and the SINR is
\begin{equation} \label{204}
\overline{SINR}=\frac{N^2\bar{\mu}_m^2 b^2\sigma_s^2} {\sigma_v^2}
\end{equation}

For the proposed scheme, the collaborating node $c_i$ transmits
$\tilde{x}_i$ given by (\ref{trans}). It can be shown the average
power of $\tilde{x}_i$ equals
\begin{eqnarray}
E\{\tilde{x}_i \tilde{x}_i^*\} = \mu_m^2\sigma_s^2 \sigma_a^4 \beta'
\end{eqnarray}
where $\beta'=K+1+\frac{\sigma_w^2}{\sigma_s^2\sigma_a^2}$.

To keep the same average transmit power as $\bar{\mu}_m^2
\sigma_s^2$,
\begin{eqnarray} \label{205}
\mu_m^2= \frac{\bar{\mu}_m^2}{\beta' \sigma_a^4}
\end{eqnarray}
Under this value of $\mu_m$ and based on (\ref{avgbeampattern}), the
average beampattern at $\phi_m$ is given by
\begin{eqnarray} \label{203}
P_{\mathrm{av}}(\phi_m)&=& \frac{N^2\bar{\mu}_m^2
b^2\sigma_s^2}{\beta'}\left(1+\frac{\beta'-1}{N} \right) +\sigma_v^2
\nonumber\\
&\stackrel{n \rightarrow \infty}{\longrightarrow}&
\frac{N^2\bar{\mu}_m^2 b^2\sigma_s^2}{\beta'} +\sigma_v^2 \ .
\end{eqnarray}

Thus, the average received power of the proposed scheme (without the
noise term $\sigma_v^2$) is $\beta'$ times less than that of
(\ref{202}). In other words, each collaborating node in the proposed
scheme needs to use $\beta'$ times more transmit power, in order to
(asymptotically) achieve the same average received power as (\ref{202}).

From (\ref{avgbeampattern}), (\ref{204}), (\ref{205}), (\ref{203}),
the SINR at the destination $d_m$ for the proposed scheme is
\begin{eqnarray} \label{SINR}
SINR &=& \frac{(1+\frac{1}{N}) N^2 \mu_m^2 b^2 \sigma_s^2
\sigma_a^4}{\frac{\beta'-2}{N}N^2 \mu_m^2 b^2 \sigma_s^2
\sigma_a^4+\sigma_v^2} \nonumber \\
&=&
\frac{(1+\frac{1}{N})\overline{SINR}}{\frac{\beta'-2}{N}\overline
{SINR}+\beta'} \stackrel{n \rightarrow \infty}{\longrightarrow}
\frac{\overline{SINR}}{\beta'}
\end{eqnarray}

Thus, compared with (\ref{204}), the average SINR of the proposed
scheme at the destination is still asymptotically scaled down by
$\beta'$.

\section{Simulations}
\label{sim}

\subsection{simulation setup}

We  investigate the performance of the above scheme
for different numbers of
collaborating nodes, i.e. $N$. For convenience, and without loss of
generality, we divide the nodes in a cluster into two pools: one
contains all potential active nodes, the total number of which
is fixed $\hat{J}$ = 32 and another contains $N$ collaborating
nodes. The directions of destinations are uniformly distributed in
$[0, 2\pi)$. The locations of collaborating nodes are uniformly
distributed within a disk with $\tilde{R}=10$ (i.e., the
radius normalized
by the wavelength). The channels among nodes in a cluster are
selected from zero-mean complex Gaussian processes, which are
constant within one slot, but vary between slots.

\subsection{BER performance}

In the proposed scheme, noise enters  the collaborating nodes, with
variance $\sigma_w^2$, and the destination, with variance
$\sigma_v^2$. Let us define $\gamma_1 \triangleq
\sigma_s^2\sigma_a^2/\sigma_w^2,$ which represents the average SNR
in the process of information sharing, and define $\gamma_2
\triangleq N^2 \mu_m^2 b^2 \sigma_s^2 \sigma_a^4/ \sigma_v^2$. Note
that $\gamma_2$ is also independent of $N$ since $\mu_m$ is of the
order of $1/N$. The overall SINR in (\ref{SINR}) can be rewritten by
\begin{eqnarray}
SINR &=&
\frac{1+\frac{1}{N}}{\frac{K-1+\gamma_1^{-1}}{N}+\gamma_2^{-1}}
\end{eqnarray}
which is determined by $\gamma_1$, $\gamma_2$, $K$ and $N$.

We fix $\gamma_1$ to 20 dB to investigate the performance of bit
error rate (BER) for different values of $\gamma_2$ and $N$. We perform a
Monte-Carlo experiment consisting of $10^6$ repeated independent
trials. Each packet contains BPSK symbols and $K = 4$ nodes are
transmitting all the time. Fig. \ref{fig1}(a) shows the
BER vs.  $\gamma_2$  for the case in which
 only one beampattern is formed in each slot, for
 different values of $N$. Solid lines correspond to estimated channels,
number of active nodes and destination information. One can see that
BER decreases as $\gamma_2$ and $N$ increase. Dashed lines
correspond to perfect knowledge of channels, number of active nodes
and destination information, and can be considered as lower bounds
of BER performance. Fig. \ref{fig1}(b) shows the BER performance in
which all of four simultaneous beams are always formed in one slot.
Compared with Fig. \ref{fig1}(a), more collaborating nodes are
needed to achieve the same BER level.

\subsection{Throughput Performance}

To investigate performance under certain traffic load $p$, we
perform a random experiment consisting of 1,000 repeated independent
trials. In each trial, all users are statistically the same, and
each user sends out packets with probability $p/\hat{J}$. The
throughput is defined as the average number of packets that were
successfully transmitted in a time slot. Each packet contains 424
bits with QPSK symbols. The nodes' ID sequences are selected based
on a $\hat{J}$th order Hadamard matrix and the IDs are used to
determine the active nodes involved in collisions as well as
to estimate the channels. A
maximum likelihood decoder is used at the destination nodes to
recover the symbols. Packets received at the destinations with BER
higher than 0.02 are considered to be lost or corrupted.
$\gamma_1=\gamma_2= 20$ dB.

In Figs. \ref{fig2} (a)-(c), we show the throughput performance
for different values of $p$ and $N$, allowing up to 1, 2, 3 simultaneous
beampatterns per slot, respectively. The curves with legend
``ideal'' correspond to the case where all the transmitted packets
are successfully received, which can be considered to be an upper bound
on the proposed scheme. One can see that the increase of $N$ can
result in throughput improvement. Furthermore, for small $N$, one
should choose a small number of simultaneous beampatterns to improve
throughput. Fig. \ref{fig2}(d) shows the throughput in which all the
$K$ beampatterns are allowed to be formed in one slot. Note that $N=1,024$
enables throughput of almost $K/2$.

\section{Conclusions and Future work} \label{con}

In this paper we have proposed a technique for reducing the time needed
for information sharing during collaborative beamforming, and for allowing
simultaneous transmissions. The proposed scheme can achieve high
throughput at the cost of reduced SINR. An analysis for the
average beampattern and network performance has also been provided.
Our analysis is based on a number of ideal assumptions on
the system. In future work, we plan to investigate the effects of
imperfect channel/phase and non-identical path loss, and also to
seek closed-form BER expressions.

\newpage
\vspace{0.5 cm}
\begin{figure}[htb]
\begin{minipage}[b]{1.0\linewidth}
 \centerline{\epsfig{figure=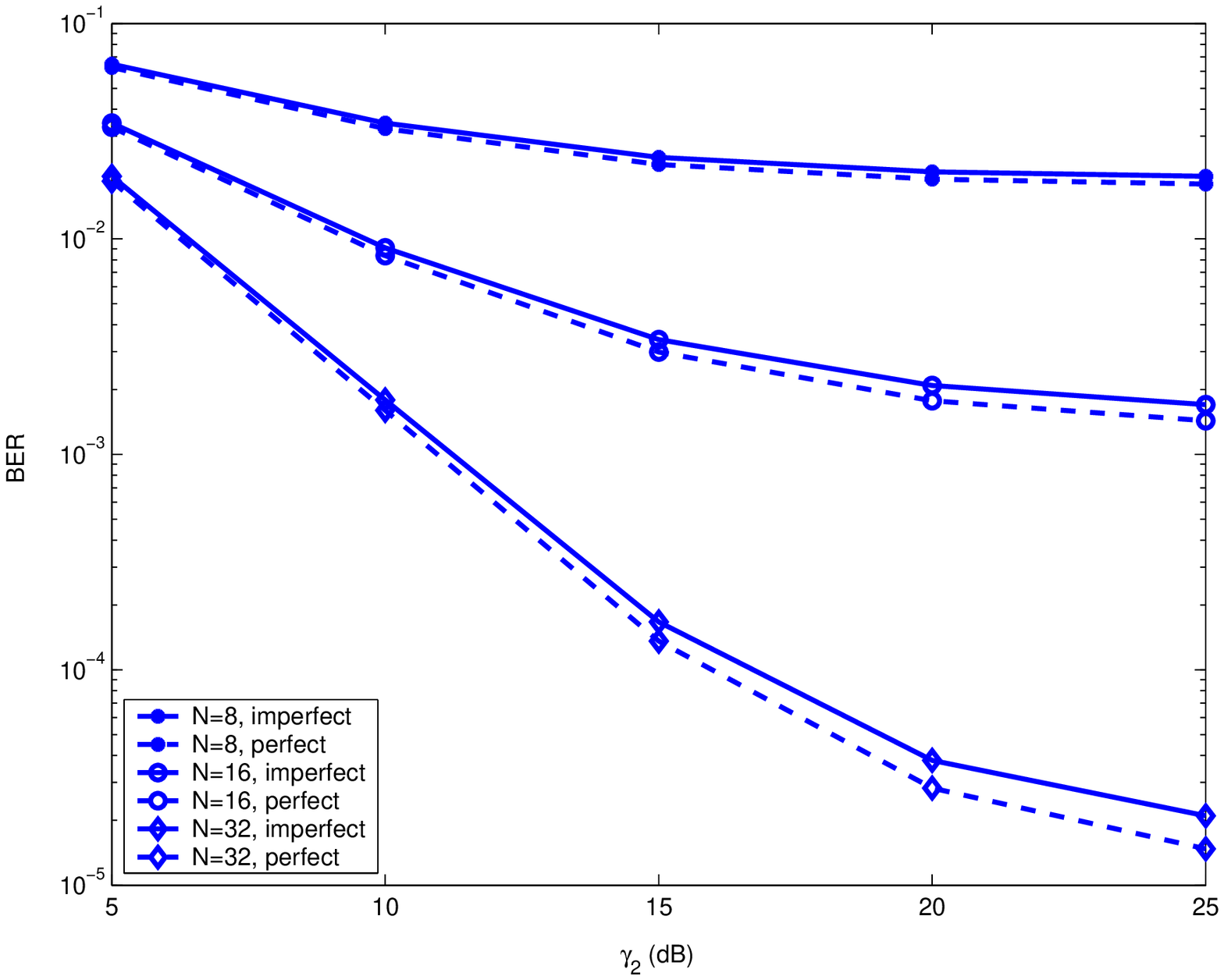,width=7.0cm}}
  \centerline{(a) one beam per slot}
\end{minipage}
\end{figure}

\begin{figure}[htb]
\begin{minipage}[b]{1.0\linewidth}
 \centerline{\epsfig{figure=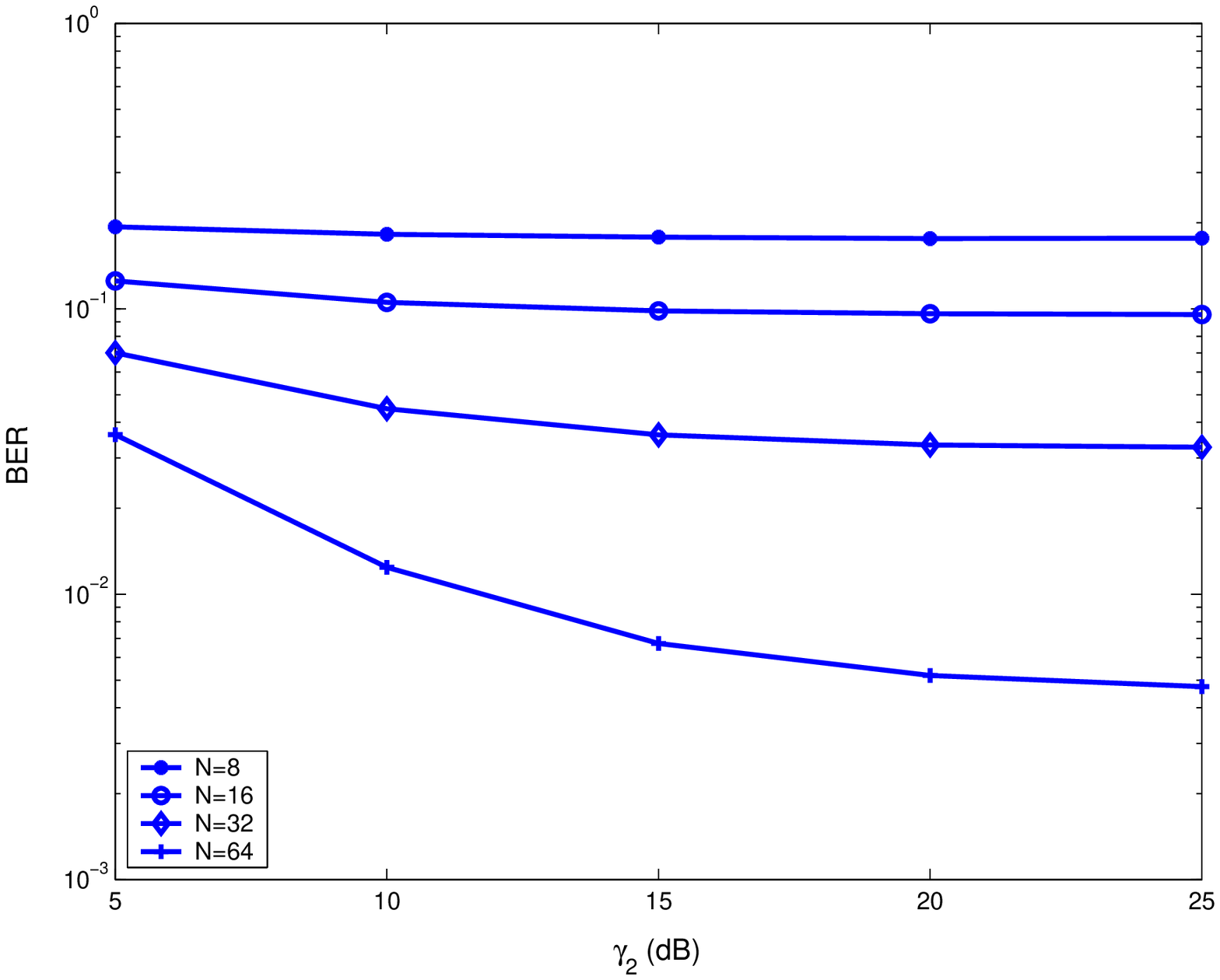,width=7.0cm}}
  \centerline{(b) 4 simultaneous beams per slot} \medskip
\end{minipage}
\caption{BER performance under different $\gamma_2$ and $N$ ($K=4$)}
 \label{fig1}
\end{figure}

\begin{figure}[htb]
\begin{minipage}[b]{1.0\linewidth}
 \centerline{\epsfig{figure=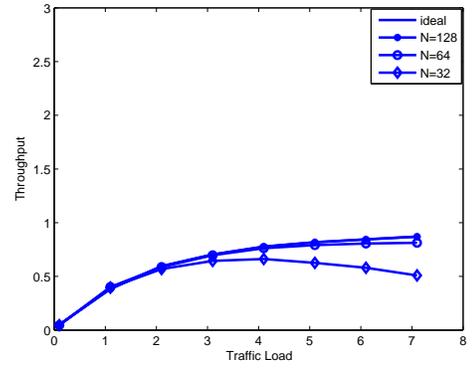,width=7.0cm}}
  \centerline{(a) up to one beam per slot}
  \vspace{0.5 cm}
\end{minipage}

\begin{minipage}[b]{1.0\linewidth}
 \centerline{\epsfig{figure=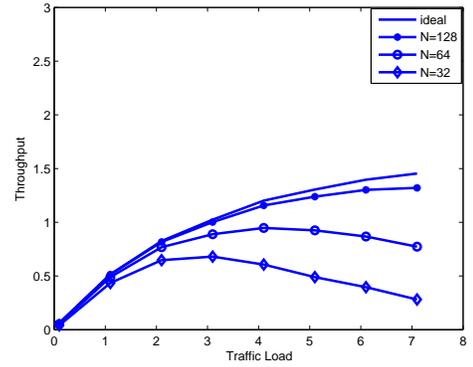,width=7.0cm}}
  \centerline{(b) up to two beams per slot}
  \vspace{0.5 cm}
\end{minipage}

\begin{minipage}[b]{1.0\linewidth}
 \centerline{\epsfig{figure=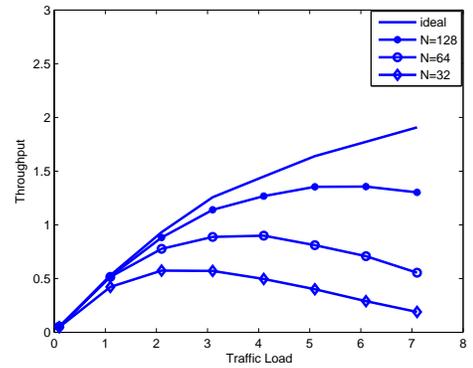,width=7.0cm}}
  \centerline{(c) up to three beams per slot}
  \vspace{0.5 cm}
\end{minipage}

\begin{minipage}[b]{1.0\linewidth}
 \centerline{\epsfig{figure=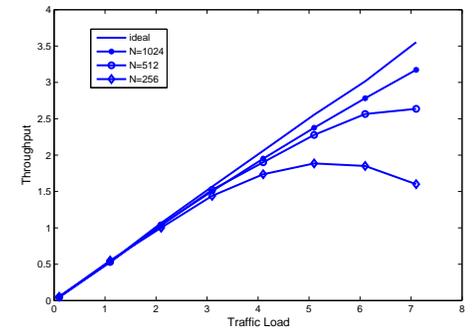,width=7.0cm}}
  \centerline{(d) up to $K$ beams per slot} \medskip
\end{minipage}
\caption{Throughput performance under different traffic load and
$N$}
 \label{fig2}
\end{figure}

\begin{thebibliography}{1}
\bibitem{Gharavi}
H. Gharavi and K. Ban, ``Multihop sensor network design for
wide-band communications,'' {\it Proceedings of the IEEE,} Vol. 91,
No. 8,  pp. 1221 - 1234, Aug. 2003.

\bibitem{alliances}
R. Lin and A. P. Petropulu, ``New wireless medium access protocol based
on cooperation,'' {\it IEEE Trans. Signal Process.}, vol. 53, no 12,
pp. 4675 - 4684, Dec. 2005.

\bibitem{Ochiai}
H. Ochiai, P. Mitran, H. V. Poor and V. Tarokh, ``Collaborative
beamforming for distributed wireless ad hoc sensor networks'', {\it
IEEE Trans. Sig. Proc.}, vol. 53,  Issue 11, pp. 4110 - 4124, Nov.
2005.

\bibitem{tse-book}
D. Tse and P. Viswanath, {\it Fundamentals of Wireless
Communication.} Cambridge University Press, Cambridge, UK, 2005.
\end{thebibliography}
\end{document}